\documentclass[useAMS,usenatbib]{mn2e}
\usepackage{aas_macros}
\usepackage{graphicx}
\graphicspath{{/misc/home/sd805/temp_figs/}}
\usepackage{fixltx2e}
\title[XTE J1739$-$302]{Discovery of the 51.47 day orbital period in the supergiant fast X-ray transient XTE J1739$-$302 with \emph{INTEGRAL}}
\author[S. P. Drave et al.]
	{S. P. Drave,$^1$\thanks{sd805@soton.ac.uk} D. J. Clark,$^{1,2}$ A. J. Bird,$^{1}$ V. A. McBride,$^{1}$ A. B. Hill,$^{3}$ V. Sguera,$^{4,6}$
        \newauthor
        S. Scaringi$^{5}$ and A. Bazzano$^{6}$ \\
	$^1$School of Physics and Astronomy, University of Southampton, University Road, Southampton, SO17 1BJ, UK \\
        $^2$Centre d'Etude Spatiale des Rayonnements, CNRS/UPS, BP 4346, 31028 Toulouse, France \\
        $^3$Laboratoire d'Astrophysique de Grenoble, UMR 5571 CNRS, Universit$\acute{e}$ Joseph Fourier, BP 53, 38041 Grenoble, France \\
        $^4$INAF-IASF, Istituto di Astrofisica Spaziale e Fisica Cosmica, Via Gobetti 101, Bologna, Italy \\
        $^5$Department of Astrophysics, IMAPP, Radboud University Nijmegen, P.O. Box 9010, 6500 GL Nijmegen, The Netherlands \\
        $^6$INAF-IASF, Istituto di Astrofisica Spaziale e Fisica Cosmica, Via del Fosso del Cavaliere 100, 00133 Roma, Italy}
\date{Accepted 2010 July 19.  Received 2010 July 9; in original form 2010 May 5}

\pagerange{\pageref{firstpage}--\pageref{lastpage}} \pubyear{2010}

\def\LaTeX{L\kern-.36em\raise.3ex\hbox{a}\kern-.15em
    T\kern-.1667em\lower.7ex\hbox{E}\kern-.125emX}

\begin{document}

\label{firstpage}

\maketitle

\begin{abstract}
Timing analysis of $\sim$12.4\,Ms of \emph{INTEGRAL}/IBIS data has revealed a period of 51.47 $\pm$ 0.02\,days in the supergiant fast X-ray transient source XTE J1739$-$302/IGR J17391$-$3021 that can be interpreted as an orbital period. An outburst history showing 35 epochs of activity has been produced, showing X-ray outbursts throughout the orbit of XTE J1739$-$302. Possible indications of an enhanced equatorial density region within the supergiant stellar wind are present in the phase-folded lightcurve. It is found that many orbital configurations are possible within this system with eccentricities of up to e $\sim$0.8 valid.
\end{abstract}

\begin{keywords}
X-rays: binaries - X-rays: individual: XTE J1739$-$302 - X-rays: bursts - stars: winds, outflows
\end{keywords}

\section{Introduction}

XTE J1739$-$302 was discovered as a new X-ray transient with the proportional counter array (PCA) on the \emph{Rossi X-ray Timing Explorer (RXTE)} \citep{XTEPCA} during observations of the Galactic centre region \citep{Smith1998}.  The source was first detected at 16:58 UT on 1997 August 12; it had not been detected in the previous scan of the region on 03 August and was not detected during the next scan on 14 August.  Assuming a thermal bremsstrahlung spectral model, \citet{Smith1998} estimated an unabsorbed peak flux (2 to 25\,keV) of 3.0~$\times$~10$^{-9}$~erg~cm$^{-2}$~s$^{-1}$.  They proposed that the source was a binary comprised of a Be star and a neutron star based upon its spectral shape but noted that its brief, yet luminous, outburst was uncharacteristic for such objects.

\citet{Sunyaev2003ATel} reported the detection of a new transient source, IGR J17391$-$3021, with the IBIS/ISGRI detector (\citealt{UbertiniIBIS}, \citealt{LebrunISGRI}) on board \emph{INTEGRAL} \citep{Winkler2003} peaking at 00:44 UT on 2003 August 26 and noting that the source position was compatible with the known location of XTE J1739$-$302.  A \emph{Chandra} localisation of the X-ray position of XTE J1739$-$302 \citep{2004ATel..218....1S} allowed the identification of the optical counterpart as an O8 Iab(f) supergiant located $\sim$2.3 kpc away \citep{2005ATel..429....1N,NegJ173912006}.

\citet{SgueraJ173912005} identified a new object class of recurrent fast X-ray transients presenting XTE J1739$-$302 as one of three example systems.  They presented four new outbursts detected by \emph{INTEGRAL} and presented a refined position which associated XTE J1739$-$302 with IGR J17391$-$3021 with a high probability.  All three systems showed the same characteristic outburst behaviour with outburst rise times of the order of an hour and typical durations of less than a day. The detection of more fast X-ray transients by \emph{INTEGRAL} and the subsequent identification of a number of the optical counterparts with OB supergiants led to the class being known as supergiant fast X-ray transients \citep[SFXTs,][]{SFXTneguer}.

This object class is characterised by quiescent or very low levels of X-ray activity with luminosity values or upper limits of 10$^{32}$ - 10$^{34}$\,erg s$^{-1}$ rising by a factor of 10$^{3}$ - 10$^{4}$ during outburst \citep{SFXTneguer}. To date there are 10 confirmed SFXTs \citep[see individual source papers for details,][]{NegJ173912006,2006A&A...455..653P, 2007A&A...469L...5R, 2008A&A...486..911N, 2008A&A...492..163R, 2008A&A...482..113M, 2009MNRAS.392...45R, 2009A&A...494.1013Z}.  There are also a number of candidate SFXTs which exhibit the appropriate X-ray flaring behaviour but for which an optical counterpart is yet to be determined \citep[e.g.][]{SFXTsguera}.

The outburst histories of some of the known SFXTs have been shown to display periodic emission, interpreted as the orbital period of the compact object in the system. These periods have a large range from a few days, such as IGR J16479$-$4514 with a period of 3.32\,days \citep{Jain2009}, to fractions of a year; IGR J11215$-$5952 having an orbital period of ~165\,days \citep{Romano2009}. In some cases an X-ray pulse period has also been discovered \citep[e.g IGR J11215$-$5952, IGR J16465$-$4507, IGR J18483$-$0311, AX J1841.0-0536,][]{2007ATel..999....1S,2006A&A...453..133W,2007A&A...467..249S,Bamba2001}, leading to the firm identification of the compact object as a neutron star. Several mechanisms to explain the observed outbursts in SFXT systems have been proposed. In general these use either a structure in the supergiant stellar wind or magnetic effects on the accretion flow. See \citet{Sidolimechrev} and the references therein for a fuller review of the current proposed outburst mechanisms.

Further analysis of outbursts from XTE J1739$-$302 found  flares lasting between 30 minutes and 3 hours and variations in absorption column density seen between bursts \citep{SmithJ173912006,SgueraJ173912005}. The X-ray spectrum of XTE J1739$-$302 has been well studied in all levels of emission (\citealt{Sidoli2008_OutOut}, \citealt{Sidoli2008_InOut}, \citealt{Romano2009b}, \citealt{Blay2008}, \citealt{Bozzo2010XMM}). XTE J1739$-$302 is often well fit with powerlaw spectra and exhibits variations in photon-index and absorption column densities between outburst and quiescence, further traits of SFXT behaviour. Broadband spectra have also shown a possible cut-off at $\sim$13\,keV and indications of absorption features at $\sim$30\,keV and $\sim$60\,keV.  These spectral shapes are characteristic of an accreting neutron star \citep{SFXTneguer}.

XTE J1739$-$302 is located approximately 2 degrees from the Galactic centre (RA 17:39:11.6, Dec -30:20:37.6 \citep{SmithJ173912006}) and as a result has a large amount of coverage from the \emph{INTEGRAL} Galactic Centre Bulge Monitoring Campaign \citep{KuulkersGBMP} and Galactic Centre Deep Exposure \citep{INTEGRALcore}. In this report new analysis undertaken on the full \emph{INTEGRAL}/IBIS data set is presented. Section 2 reports the data selection used and Section 3 the analysis performed, in the form of periodicity analysis in Section 3.1 and outburst identification in Section 3.2. A discussion of the results and their physical interpretation is given in Section 4, followed by conclusions in Section 5.

\section{Data Selection}

Our analysis used data from \emph{INTEGRAL} archives covering MJD\,52671.7 through MJD\,54763.6, an initial data set of $\sim$14.4\,Ms. Using the \emph{INTEGRAL} Offline Science Analysis (OSA; \citealt{GoldwurmOSA}) v7.0 software an 18 - 60\,keV Science Window (ScW) IBIS/ISGRI lightcurve was generated \citep{TonyCat4}. Nominally a ScW represents an $\sim$2000\,s observation. The lightcurve was filtered as to disregard ScWs with exposure times of less than 200\,s and/or source locations separated from the telescope pointing axis by more than 12 degrees. As a result, ScWs with large errors on the flux determination are removed, which is desirable when data is to be used for Lomb-Scargle analysis, see Section 3.1. The result is an 18 - 60\,keV lightcurve, spanning MJD\,52698.2 to MJD\,54763.6 with a total effective exposure of $\sim$12.4\,Ms. 

XTE J1739$-$302 also has coverage in the \emph{RXTE} All Sky Monitor (ASM) and \emph{Swift} Burst Alert Telescope (BAT) public databases. After filtering for poor quality data flags, short exposures and low coded aperture fractions, the \emph{Swift}/BAT lightcurve covers MJD\,53413 through to MJD\,55208 with an effective exposure time of $\sim$13.2\,Ms. The \emph{RXTE}/ASM lightcurve spans MJD\,50088.1 to MJD\,55209.9, with a total exposure time of $\sim$3.45\,Ms.

\section{Data Analysis}

\subsection{Periodicity Analysis}

\begin{figure}
 \includegraphics[scale=0.5]{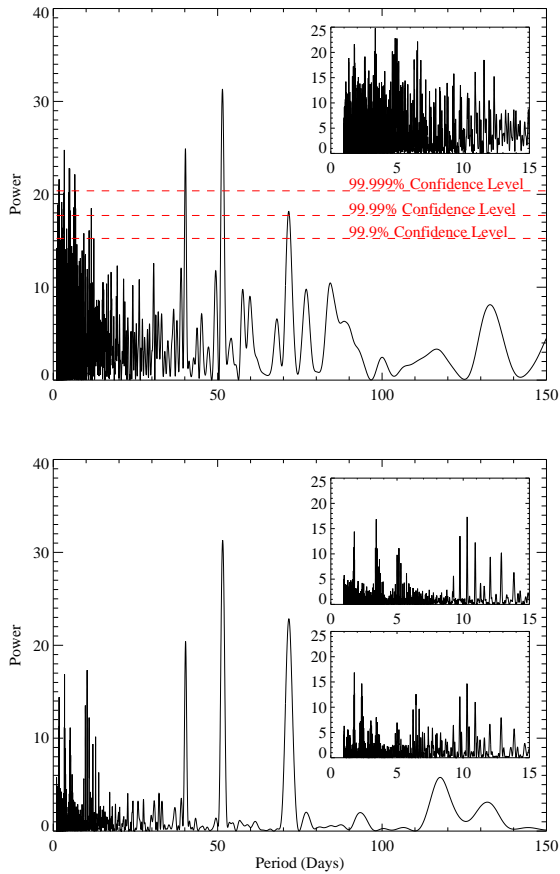}
 \caption{Top: Lomb-Scargle periodogram of the full IBIS/ISGRI 18 - 60\,keV lightcurve with the 99.9\%, 99.99\% and 99.999\% confidence levels shown. Bottom: Lomb-Scargle periodogram of a function based on the shape of the phase folded lightcurve at a period of 51.47 days modulated with the data gaps present in the IBIS/ISGRI data. Inset are zoomed views of the short period regions of the periodogram in both cases and also of a datasubset in the bottomost inset} 
 \label{fig1}
\end{figure}

The filtered IBIS/ISGRI lightcurve was tested for periodic signals by means of a Lomb-Scargle analysis (\citealt{LOMB1976}, \citealt{SCARGLE1982}). The resulting periodogram contained several peaks of relatively high power. To test for significance a Monte-Carlo based randomisation test was performed, as outlined in \citet{AdamMCT}. The resulting periodogram and calculated confidence levels are shown in Fig. \ref{fig1}.

\begin{figure}
 \includegraphics[scale=0.5]{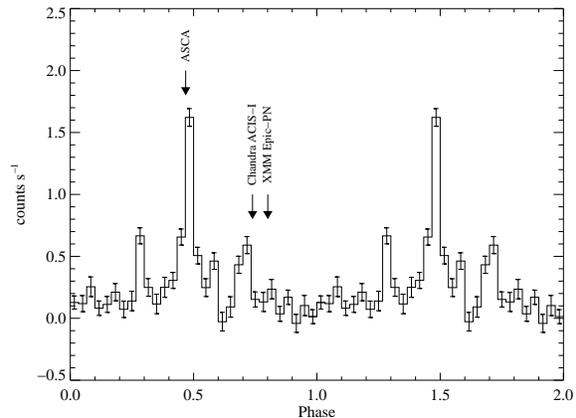}
 \caption{Phase folded lightcurve of the full IBIS/ISGRI XTE J1739$-$302 18 - 60\,keV data set with the ephemeris MJD\,52698.2 and a period of 51.47\,days. The positions of the low and quiescence level emission detections described in Section 4 are indicated.}
 \label{fig2}
\end{figure}

As Fig. \ref{fig1} shows, there are several peaks that appear to be above the 99.999\% significance level. However, as a binary system containing this number of physical periodicities is unlikely, efforts were made to explain the presence of each signal. The first effect considered was that of the window function resulting from the large scale, non-uniform sampling of data within the IBIS/ISGRI lightcurve. To evaluate these effects, simulated lightcurves with the same shape as the phase-folded lightcurves (e.g. Fig. \ref{fig2}) and identical data gaps to the real data were generated, subjected to Lomb-Scargle analysis and their periodograms produced. Only a period of 51.47\,days reproduced patterns seen in the actual periodogram; this being also the period with the maximum power in the data. As seen in the lower panel of Fig. \ref{fig1} the periodogram of this window function explains the large features at $\sim$40 and 70\,days as well as features at shorter periods. These short period features are consistent with many of the periods above 99.999\% confidence in the real data and are interpreted as higher frequency components introduced by the non-sinusoidal profile of the phase-folded lightcurve.

While the low period peaks do not entirely match these seen in the real periodogram, this analysis was only performed for the time-averaged phase-folded lightcurve. The analysis was repeated for three subsets of the data, and the peaks at short periods were found to vary between subsets (e.g. the lowest inset in Fig. \ref{fig1}), most likely due to changes in observing pattern and frequency during the mission. Furthermore there are other possible sources of periodicity within this period range. The \emph{INTEGRAL} orbital period is $\sim$3 days and during the core program \citep{INTEGRALcore} \emph{INTEGRAL} was performing scans of the galactic plane and galactic centre exposures in a very regimented manner. As a result the shorter period significant signals are taken as being either directly related to the 51.47\,day signal, from other sources of artificial periodicity or beating interactions between these. Therefore 51.47 days is considered to be a real, physical periodicity in the XTE J1739$-$302 system. On the contrary no indication of the $\sim$8\,day period predicted by \citet{Blay2008} is found in the analysis.

A Monte-Carlo based test was also used to estimate the uncertainty on the period found within the data. Flux values are randomised within their 1$\sigma$ error bars and a Lomb-Scargle test performed, the period with highest power in the region around the initial period is recorded. This process is repeated 200,000 times and the resulting distribution fit with a Gaussian, the width of which is taken as the 1$\sigma$ error on the identified period. An error of $\pm$ 0.02\,days was found for the 51.47\,day period of XTE J1739$-$302. Figure \ref{fig2} shows the phase-folded lightcurve of the XTE J1739$-$302 system using the identified period. The ephemeris of MJD\,52698.2 is used, defining the initial zero phase at the first observation of XTE J1739$-$302 with \emph{INTEGRAL}/IBIS. The phase-folded lightcurve shows a shape that is dominated by enhancement of emission at what appears to be three points in the orbit. The supposed periastron, phase 0.4 to 0.6, also has two nearly symmetric side-peaks at phases of 0.25 to 0.35 and 0.65 to 0.75. A more detailed discussion of this profile is given in Section 4.

The public access \emph{RXTE}/ASM and \emph{Swift}/BAT lightcurves were also searched for periodic signals using the Lomb-Scargle method, however in these cases no significant signals were detected at any period. While this is unusual it is not the first time that periods have been seen with one instrument and not others \citep{DaveJ17544}. 

\subsection{Outburst Identification}

In conjunction with periodicity analysis the IBIS/ISGRI lightcurve was also searched for outbursts by a significance test. The lightcurve is searched with windows of increasing size in all positions along its length and the significance across the entire data window calculated. The data window with the maximum significance is then recorded and those ScWs removed from the lightcurve. This process is repeated until all regions with a significance greater than 4$\sigma$ have been identified. 

\begin{figure}
 \includegraphics[scale=0.5]{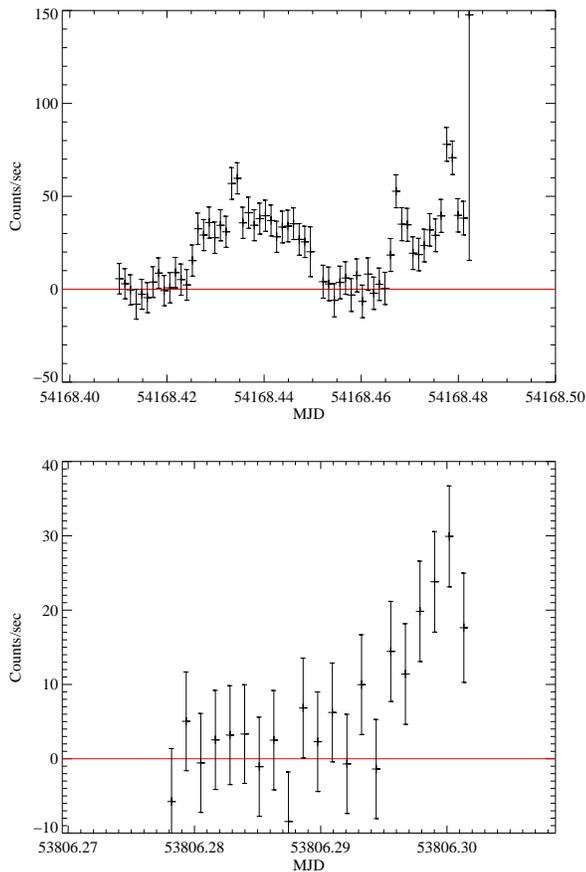}
 \caption{Finer binned 100s outbursts showing temporal behaviour at different levels of activity. Top: the 35.29$\sigma$ outburst recorded in Table 1. Bottom: a 5.42$\sigma$ outburst recorded at MJD\,53806.29. The larger outburst displays multiple flares within regions of consistent emission, whereas the smaller outburst consists of one lower level flare.}
 \label{fig3}
\end{figure}

Statistical tests were performed to estimate how many detections could result from a combination of random noise in the data and the number of trials performed. A synthetic random lightcurve was created with the same statistical properties as the real data (RMS and number of data points) and run through the burst finding procedure. This resulted in five `detections' above the 4$\sigma$ level, the largest being 4.41$\sigma$. Consequently, the seven events below 4.5$\sigma$ were removed from the outburst history. Finally higher time resolution lightcurves (100\,s binning) were generated for each of the remaining events and checked manually to ensure outburst behaviour could be observed. It was seen that the majority of the remaining events showed distinct flaring features while the remainder showed lower level activity, possibly due to small flares with emission mostly below the sensitivity of IBIS/ISGRI but where the peak of the emission is detected. As a result thirty five outburst events have been identified within the XTE J1739$-$302 IBIS/ISGRI lightcurve. The subset of all those outbursts with a significance greater than 8$\sigma$ is given in Table 1, and examples of finely binned outburst lightcurves at high and low levels of activity are shown in Fig \ref{fig3}.

\begin{figure}
 \includegraphics[scale=0.5]{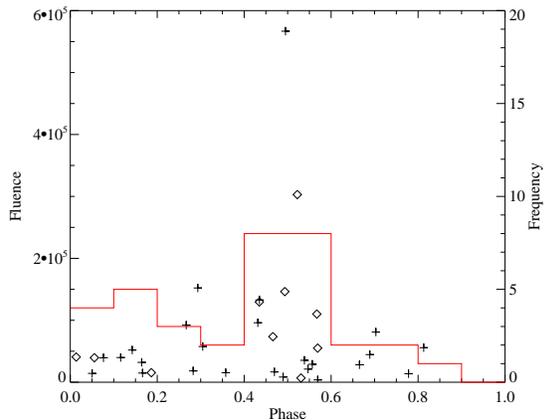}
 \caption{Outburst distribution in orbital phase space. The histogram shows the number distribution of outbursts in orbital phase (right hand scale) whereas the individual points show fluence estimations, in counts, of each outburst (left hand scale, crosses are full estimates and diamonds lower limits). While the number distribution appears somewhat flat, it is seen that the fluence measurements are peaked around periastron. Hence while outbursts can occur at any point around the orbit, the larger outbursts are observed at periastron.}
 \label{fig4}
\end{figure}

Combining the outburst history with the new orbital ephemeris allows the phase of each outburst event to be defined. Figure \ref{fig4} shows the phase distribution of the identified outbursts. The histogram shows the number of outbursts observed around the orbit (right hand axis) and it is seen that the distribution is not strongly structured. However, using the finely binned lightcurves of each event, a measurement of the fluence was made and this distribution is shown by the points in Fig. \ref{fig4}. Crosses indicate measurements where a full outburst profile was deemed to be present, whereas diamonds represent lower limits determined from profiles that showed incomplete coverage of the event. It can be seen that the `periastron' region contains more larger outbursts. Hence while it appears possible to observe outbursts at any point around the orbit of XTE J1739$-$302, the larger outbursts occur predominantly at periastron. A cluster of higher fluence outbursts are also seen within the phase of the leftmost side-peak of Fig. \ref{fig2}. Targeted pulsation searches were performed on the most significant of the identified outburst ScWs, however no significant pulsations were detected.

\begin{figure}
 \includegraphics[scale=0.5]{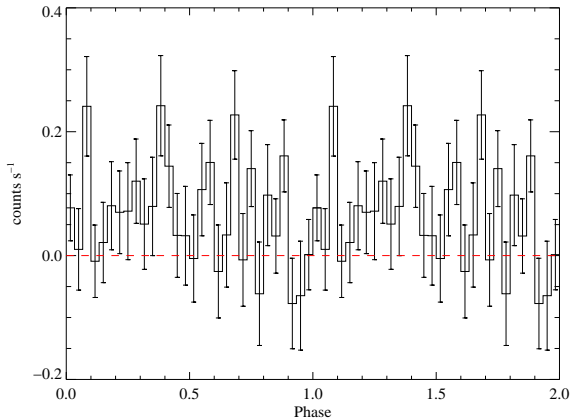}
 \caption{Phase folded 18-60 keV lightcurve with outbursts removed, folded on the 51.47 day period. There appears to be some consistently non-zero, low level emission around the orbit, however it is not periodic in nature, the maximum power generated in the Lomb-Scargle analysis was 13 and not significant. Hence it is believed that for this system the periodic signal is generated purely from outburst emission.}
 \label{fig5}
\end{figure}

The IBIS/ISGRI lightcurve was also analysed after the outbursts were removed to evaluate if any underlying flux modulation could be observed. The folded lightcurve with the outbursts removed (Fig. \ref{fig5}) shows no sign of an underlying periodicity, suggesting that the 51.47 day signal observed in the periodogram is a direct result of outburst emission. However there does appear to be some very low level detectable emission within the orbit, comparable to the 2 - 10\,keV flux detected by \emph{Chandra} of 1.3 $\times$ 10 $^{-11}$\,erg cm$^{-2}$ s$^{-1}$ (using the spectral parameters $\Gamma$ = 0.62 and n$_{H}$ = 4.2 $\times$ 10$^{22}$\,cm$^{-2}$ of \citealt{SmithJ173912006}) occurring at a phase of 0.793 with these ephemeris. The location of this \emph{Chandra} ACIS-I observation is indicated in Fig. \ref{fig2}.

The \emph{Swift}/BAT orbital lightcurve was also inspected for outbursts and it was seen to contain twelve outburst events. Coincident detections with IBIS/ISGRI are indicated in Table 1. There were also six events that are not temporally coverred by the IBIS/ISGRI data, these are outlined in Table 2. Of the whole set five outbursts occurred in the periastron region, four within the `side peaks' and three at other phase locations, showing a phase distribution similar to that seen in the IBIS/ISGRI lightcurve, see the phase ratios in Section 4. As the periodic signal in this source is believed to come mostly from outburst emission the lower number of events in the \emph{Swift}/BAT lightcurve helps explain why the signal is not seen in the periodicity analysis in this case.

\begin{table*}
\begin{minipage}{150mm}
\caption{IBIS/ISGRI identified outbursts with a significance of greater than 8$\sigma$}
\begin{center}
\begin{tabular}{|c|c|c|c|c|c|c|}
\hline
\multicolumn{1}{|c|}{Date MJD$^{a}$} & \multicolumn{1}{c|}{Significance$^{b}$} & \multicolumn{1}{c|}{Orbital Phase} & \multicolumn{1}{c|}{Peak Flux$^{b}$} & \multicolumn{1}{c|}{Luminosity$^{b,c}$} & \multicolumn{1}{c|}{Duration} & \multicolumn{1}{c|}{Also Reported} \\
- & $\sigma$ & - & 10$^{-10}$ erg cm$^{-2}$ s$^{-1}$ & 10$^{35}$ erg s$^{-1}$ & (h) & In \\ \hline
52720.52 & 36.17 & 0.44 & 19.03 & 12.04 & 1.55 & \citealt{SgueraJ173912005} \\
52724.63 & 8.17 & 0.55 & 11.34 & 7.18 & 41.20 & ** \\
52877.8 & 49.22 & 0.5 & 22.90 & 14.50 & 14.59 & \citealt{Sunyaev2003ATel} \\
52892.62 & 8.31 & 0.78 & 5.55 & 3.51 & 0.38 & ** \\
53073.51$^{d}$ & 35.91 & 0.29 & 16.20 & 10.26 & 2.31 & \citealt{SgueraJ173912005} \\
53074.09$^{d}$ & 16.64 & 0.3 & 10.03 & 6.35 & 1.26 & \citealt{SgueraJ173912005} \\
53093.86 & 12.72 & 0.69 & 5.43 & 3.44 & 1.75 & ** \\
53238.24 & 39.42 & 0.49 & 26.46 & 16.75 & 1.82 & \citealt{SFXTsguera} \\
53248.97 & 26.48 & 0.7 & 13.04 & 8.26 & 1.71 & ** \\
53479.78 & 13.29 & 0.19 & 11.41 & 7.22 & 0.33 & \citealt{SmithJ173912006} \\
53802.97$^{e}$ & 27.5 & 0.47 & 17.20 & 10.88 & 0.63 & ** \\
53831.15 & 14.33 & 0.01 & 8.69 & 5.50 & 0.64 & ** \\
53987.7$^{e}$ & 13.76 & 0.06 & 8.25 & 5.22 & 0.62 & ** \\
53990.81 & 14.77 & 0.12 & 7.70 & 4.87 & 1.25 & ** \\
54011.69$^{e}$ & 28.4 & 0.52 & 15.22 & 9.63 & 11.60 & ** \\
54161.08$^{d,e}$ & 10.26 & 0.43 & 4.69 & 2.97 & 9.00 & ** \\
54161.57$^{d,e}$ & 26.7 & 0.43 & 24.67 & 15.62 & 1.44 & \citealt{Turler2007ATel} \\
54168.43$^{e}$ & 35.29 & 0.57 & 16.27 & 10.30 & 1.37 & ** \\
54564.68$^{e}$ & 23.8 & 0.27 & 14.634 & 9.26 & 0.38 & \citealt{Romano2008ATel} \\ \hline
\end{tabular}
\end{center}
\medskip
 Note: The peak flux represents the highest count rate achieved in a ScW during each outburst. These are converted in to the given flux with the conversion 1.3$\times$10$^{-11}$\,erg cm$^{-2}$ s$^{-1}$ = 1.4$\times$10$^{-4}$\,ph cm$^{-2}$ s$^{-1}$. $^{a}$ MJD is that of the mid-point of the first ScW identified in an outburst event. $^{b}$ Energy range 18 - 60\,keV. $^{c}$ Assuming a distance of 2.3\,kpc. $^{d}$ Part of the same outburst event. $^{e}$ Detection in the \emph{Swift}/BAT orbital lightcurve
.
\end{minipage}
\label{table1}
\end{table*}

\begin{table}
\caption{\emph{Swift}/BAT outbursts not temporally coverred by IBIS/ISGRI}
\begin{tabular}{|c|c|c|c|}
\hline
\multicolumn{1}{|c|}{Date MJD} & \multicolumn{1}{c|}{Phase} & \multicolumn{1}{c|}{Significance} & \multicolumn{1}{c|}{Duration (h)} \\ \hline
53424.69 & 0.10 & 8.91 & 2.4 \\
53765.79 & 0.73 & 12.64 & 18.3 \\
54269.14 & 0.52 & 7.53 & 3.2 \\
54347.45 & 0.05 & 7.75 & 6.2 \\
54411.19 & 0.28 & 8.29 & 3.1 \\
54692.00 & 0.74 & 17.47 & 11.2 \\ \hline
\end{tabular}
\label{table2}
\end{table}

\section{Discussion}

An orbital period of 51.47\,days puts XTE J1739$-$302 as one of the longer period SFXTs, approximately one third the length of the longest SFXT, IGR J11215--5952 with its period of 165\,days. The large size of this orbit gives rise to many possible orbital configurations, making the constraining of orbital parameters difficult. Using Kepler's third law, assuming a neutron star mass of 1.4\,M$_\odot$ and a stellar mass range of 25 - 28\,M$_\odot$ for the supergiant \citep{AllenPQ}, the semi-major axis of the system is estimated to be 173.4 - 179.8\,R$_\odot$. The lower and upper limits of the stellar radius are estimated as 14.4 - 23.2\,R$_\odot$ from the SED fitting outlined in \citet{Rahoui2008}. For the remainder of this discussion we adopt a stellar mass of 25\,M$_\odot$ and an orbital semi-major axis of 173.4\,R$_\odot$.

\begin{figure}
 \includegraphics[scale=0.5]{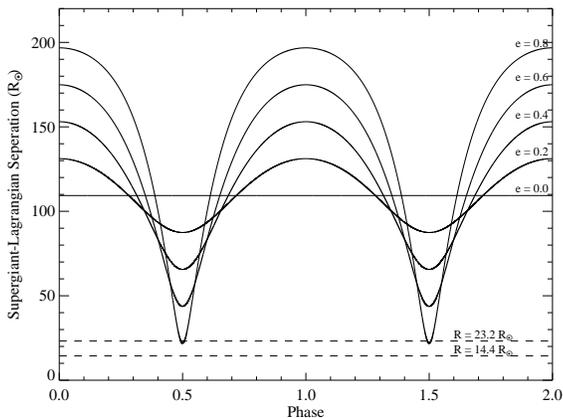}
 \caption{The L1 Lagrangian point separation from the Supergiant as a function of orbital phase. The eccentricities 0.0, 0.2, 0.4, 0.6 and 0.8 are plotted. Eccentricities of up to e $\sim$0.8 are possible before a Roche-Lobe overflow regime would start to take effect at the upper limit of the supergiant radius.}
 \label{fig6}
\end{figure}

By inspecting the phase-folded lightcurve, Fig. \ref{fig2}, it can be seen that with this ephemeris the maximum flux occurs in the orbital phase bin centred on 0.483. Further inspection of this region of the orbit, from Table 1, shows variations in the properties of outbursts. The fact that an outburst is not observed with each periastron passage and that the duration of observed outbursts in this region vary from 0.63 to 41.2\,hours indicates that the XTE J1739$-$302 system does not undergo accretion in the Roche-Lobe overflow regime, whereby persistent periastron emission on a more consistent timescale would be expected. Figure \ref{fig6} shows a plot of the L1 Lagrangian point separation from the supergiant companion as a function of orbital phase for a range of orbital eccentricities \citep{Paczynski1971}. As this shows, eccentricities of up to $\sim$0.8 are possible before the separation moves to within the upper limit of the stellar radius, inducing Roche-Lobe overflow. Therefore from this consideration alone it is not possible to place a tightly constraining limit on the eccentricity of the system. 

The low luminosity state of 8.2 $\times$ 10$^{33}$\,erg s$^{-1}$ (2 - 10 keV), assuming a distance of 2.3 kpc, observed with \emph{Chandra} \citep{SmithJ173912006} and the quiescence detections of a 3$\sigma$ upper limit at $<$7.0 $\times$ 10$^{32}$\,erg s$^{-1}$ (2 - 10 keV), 6.0 $\times$ 10$^{32}$\,erg s$^{-1}$ (2 - 10 keV) and 4.1 $\times$ 10$^{32}$\,erg s$^{-1}$ (0.5 - 10 keV) by \emph{ASCA} \citep{Sakano2002}, \emph{Swift}/XRT \citep{Romano2009b} and \emph{XMM}/Epic-PN \citep{Bozzo2010XMM} respectively are not consistent with spherically symmetric, smooth wind Bondi-Hoyle accretion \citep{BondiHoyle1944}. While the variations in emission could result from NS separation changes around the orbit for some of the possible orbital configurations in this system, the lack of an underlying flux modulation in Fig. \ref{fig5} and the inconsistency in significance of different outbursts at approximately the same phase of the orbit shows that a smoothly varying stellar wind can not explain the outbursts observed. Instead we take this to show that the neutron star is in fact travelling through an inhomogeneous, non-symmetrical stellar wind in the XTE J1739$-$302 system, in accordance with the ``clumpy wind models'' of SFXTs (\citealt{2007A&A...476..335W}, \citealt{2008AIPC.1010..252N}).

Looking at Fig. \ref{fig2} it can be seen that the \emph{Chandra} observation occurred at the edge of one of the emission side peaks. Hence the higher level of emission seen in that data compared to the `quiescent data' could be resulting from observation during an orbital phase that shows a systematic increase in luminosity over the full IBIS/ISGRI lightcurve. Figure \ref{fig5} shows the phase folded lightcurve of XTE J1739$-$302 with all identified outbursts removed, we see that there is residual non-zero emission remaining in this lightcurve. Taking the average count rates out of this data a luminosity of 2.9 $\times$ 10$^{33}$\,erg s$^{-1}$ (18 - 60\,keV) is calculated, using the conversions outlined in Table 1. The fact that persistent emission is observed in the full out-of-outburst IBIS/ISGRI lightcurve, shows some underlying activity in the system. Following the arguments in the above paragraph this activity is attributed to many small flares that can not be detected individually but sum up to a detectable emission over the full length of the IBIS/ISGRI lightcurve. Behaviour such as this was observed in the \emph{XMM} observations of \citet{Bozzo2010XMM} and supports the conclusions of \citet{Romano2009b} that `true quiescence is a rare state'. This is a different behaviour to that observed in IGR J17544$-$2619 \citep{DaveJ17544} where the removal of outbursts resulted in a relatively small change in the shape of the lightcurve. The presence of underlying periodic modulation of the emission in that system and the lack of it in this one is likely an orbital effect. The larger orbit of XTE J1739$-$302 prohibits observable modulation being produced by the orbital variations in either a smooth stellar wind component or the frequency of low level flaring events. XTE J1739$-$302 shows the extent of its transient nature from the \emph{ASCA} observation, where a deep quiescence was observed followed by rapid flares with dynamic ranges of over 10$^{3}$, and took place at a phase of 0.468 in this ephemeris, placing it within the periastron region. This suggests a very high degree of `clumping' within the wind in order to allow quiescence states, in very close proximity to distinct outbursts, to be observed within the periastron region of the orbit. This is taken to show that the majority of the periodicity information in XTE J1739-302 is a result of outburst emission. To safeguard against a small number of events separated by multiples of the proposed orbital period falsely creating a signal, iterative and random removals of identified outbursts from the lightcurve were performed. There were no indications from either case that the removal of a small number of events could destroy the signal. It is believed that the reliance on outburst detection contributes to the periodicity only being seen by IBIS/ISGRI as a combination of instrument sensitivities and pointing strategies could prevent outburst detection in other instruments.

The presence of the two side-peaks seen in Fig. \ref{fig2}, both detected at $\sim$4 to 5$\sigma$ significance is intriguing and could help in defining the orbital characteristics of the system. In \citet{Ducci2009} an anisotropic stellar wind with an enhanced equatorial density region, inclined at some angle to the plane of the neutron star orbit, is invoked within the clumpy wind model to explain features in the phase-folded lightcurve of IGR J11215--5952. As a consequence a more general system was modelled and it was shown that symmetric outbursts corresponding to the crossing of the enhanced density region by the neutron star could produce up to 3 outbursts per orbit (Figure 15 of \citealt{Ducci2009}). Using a simplified geometric version of this model with the equatorial density region inclined to the neutron star orbit at 90$^{o}$, intersections of the neutron star orbit and stellar wind disc at the required phases are obtained for an orbit with an eccentricity of e $\sim$0.16. However this results in a difference in separation between periastron and apastron of 56.3\,R$_{\odot}$, corresponding to a ratio of clump interaction probability of only P$_{peri}$ $\simeq$ 1.5\,P$_{ap}$ using the relationship in \citet{DaveJ17544}. This is below the observed distribution of outbursts, N$_{peri}$ = 1.89\,N$_{ap}$ for all outbursts or N$_{peri}$ = 2.25\,N$_{ap}$ for just the larger outburst events detailed in Table 1. This increase is to be expected in this model as clumps expand as they move out from the parent star, hence becoming less dense and making strong outbursts less likely. Since 90$^{0}$ is the most extreme case of the inclination of the enhanced density region to the orbital plane, however, the value of e $\sim$0.16 can act as a lower limit on the eccentricity if the enhanced equatorial density region explanation proves to be correct. 

If the side-peaks are a real feature then this also has wide implications for the interpretation of SFXT systems. The presence of equatorial disc-like structure within the stellar wind goes beyond the current applications of orbital variations, compared to more classical Sg-XRBs, to explain the observed behaviour of SFXTs and implies a link towards the Be-XRB class of systems as well. This would be strong evidence to show that SFXTs are in fact an intermediate class of systems that help bridge the gap between the two original classes of HMXBs.

\section{Conclusions}

Observations of XTE J1739$-$302 across the lifetime of the \emph{INTEGRAL} mission using IBIS/ISGRI have uncovered a period of 51.47 $\pm$ 0.02\,days which is interpreted as the orbital period of the system. The relatively long period, in SFXT terms, results in an orbital semi-major axis in the range 173.4 - 179.8\,R$_{\odot}$ and allows for orbital eccentricities up to $\sim$0.8 before Roche-Lobe overflow emission would be expected. No further, firm constraints can be placed on the eccentricity at the current time. The outburst history and reported luminosities lead to the conclusion that the neutron star orbits in an inhomogeneous `clumped' stellar wind, with some evidence that this system may contain an enhanced equatorial density region. The fact that emission at quiescent levels can be seen very near periastron suggests a high degree of `clumping' within the stellar wind. Further observations and modelling are encouraged to identify with greater certainty if features in the phase-folded lightcurve are in fact due to an enhanced equatorial density region, as well as directed optical follow-up in an attempt to identify the presence or lack of any features that are akin to the Be X-ray binaries that contain prominent equatorial discs. The identification of an orbital period in XTE J1739-302 makes it the sixth firmly identified periodic SFXT, suggesting that periodicity is a common feature in such systems. It also gives a further behaviour to a system that is widely considered the archetype of the SFXT class.

\section*{Acknowledgements}
Based on observations with \emph{INTEGRAL}, an ESA project funded by member states (especially the PI countries: Denmark, France, Germany, Italy, Switzerland, Spain), Czech Republic and Poland, and with the participation of Russia and the U.S.A. S. P. Drave and V. A. McBride acknowledge support from the Science and Technology Facilities Council, STFC. A.~B. Hill acknowledges funding by contract ERC-StG-200911 from the European Community. A. Bazzano acknowledges support through the contract ASI-INAF 008/07. V. Sguera acknowledges financial support via grant ASI-INAF /88/0//and /08/0//.

%\bibliographystyle{mn2e}
%\bibliography{IGRJ17391}

\label{lastpage}

\end{document}